\documentclass[aps,prd,twocolumn,showpacs,amsmath,amssymb,nofootinbib, showpacs, showkeys]{revtex4-2}
\usepackage{graphicx}
\usepackage{color}
\usepackage{times}
\usepackage{inputenc}
\usepackage{bm}
\usepackage{ulem}
\usepackage{multirow}
\usepackage{float}
\usepackage{url}
\usepackage{mathrsfs}
\usepackage{physics}
\usepackage{comment}
\usepackage[table,xcdraw]{xcolor}
\usepackage{booktabs,makecell}
\usepackage{comment}
\usepackage[colorlinks=true,citecolor=blue,urlcolor=blue,linkcolor=blue]{hyperref}
\usepackage[caption=false]{subfig}

\usepackage{booktabs}
\usepackage{multirow}

\usepackage{graphicx}
\usepackage{dcolumn}
\usepackage{bm}

\begin{document}

\preprint{APS/123-QED}

\title{Nuclear Pasta and Crustal Quasi-Periodic Oscillations in Neutron Star }
\author{Vishal Parmar$^{1}$}
\email{vishal.parmar@pi.infn.it}
\author{Ignazio Bombaci$^{1,2}$}
\affiliation{\it $^{1}$ INFN, Sezione di Pisa, Largo B. Pontecorvo 3, I-56127 Pisa, Italy}
\affiliation{\it $^{2}$ Dipartimento di Fisica, Universit\`{a} di Pisa, Largo B.  Pontecorvo, 3 I-56127 Pisa, Italy}

\date{\today}

\begin{abstract}
We investigate the impact of nuclear pasta on crustal structure and
torsional oscillations using a Bayesian ensemble of unified neutron-star equations of
state based on relativistic mean-field models constrained by nuclear experiments,
empirical saturation properties, chiral effective field theory, and
multimessenger observations. For each posterior sample, we compute the pasta sequence
within a compressible liquid-drop model and quantify the onset density, thickness,
and mass fraction of the pasta layers. We show that the appearance and extent of
nuclear pasta are primarily controlled by the  symmetry-energy slope parameter $L$. While
spherical and rod-like pasta configurations are present for all equations of state,
only a small fraction of the posterior supports slab, tube, or bubble geometries. The
transition from spherical nuclei to rods is tightly constrained to occur at a density
of $\rho_{\rm sr} = 0.0588^{+0.0045}_{-0.0065}\,\mathrm{fm^{-3}}$. We further predict that the nuclear
pasta layer occupies a relative radial thickness of
$\Delta R_{\rm pasta}/\Delta R_{\rm c} = 0.140^{+0.025}_{-0.036}$ and contributes a
relative mass fraction of
$\Delta M_{\rm pasta}/\Delta M_{\rm c} = 0.475^{+0.071}_{-0.113}$.  Using the resulting crust models, we present the first quasi-periodic oscillations (QPOs) analysis based on a Bayesian
posterior ensemble of neutron-star equations of state and systematically assess their
compatibility with observed low-frequency quasi-periodic oscillations. We find that
the predicted QPO frequencies are strongly correlated with the curvature of the
symmetry energy evaluated at sub-saturation density, $K_{\rm sym}(\rho_0/2)$, and that
uncertainties in the equation of state translate into a range of angular indices
$\ell$ consistent with the observed frequencies. 
\end{abstract}

\maketitle

\section{\label{intro} Introduction}
Although the neutron-star crust comprises only $\sim$10\% of the stellar radius, it plays a central role in several key observables, including neutron-star mergers, magnetar quasi-periodic oscillations, thermal relaxation, and pulsar glitches \cite{Haensel_2008}. The outer crust consists of an elastic solid, body-centred cubic Coulomb lattice of neutron-rich nuclei embedded in a degenerate electron gas, while neutron drip marks the transition to the inner crust, where nuclei coexist with a neutron superfluid \cite{Gearheart_2011}. Near the crust–core boundary, the competition between nuclear  and  Coulomb repulsion leads to geometrical frustration, giving rise to non-spherical nuclear configurations collectively known as nuclear ``pasta'' \cite{Caplan_2018}. While the pasta layer themselves occupies only $\sim$10\% of the total crust thickness, it can significantly alter the elastic and transport properties of the crust \cite{rezzolla2018physics} and play a crucial role in influencing various mechanisms such as, crust cooling \cite{Horowitz_2015, Ootes_2018}, spin period \cite{Pons_2013}, quasiperiodic oscillation in giant flares \cite{SteinerWatts2009PRL103}, shattering of the crust \cite{Tsang_2012}. While no direct and robust observational evidence for nuclear pasta currently exists, several indirect and suggestive observational signatures have been proposed \cite{Horowitz_2014,  Pons_2013}. Consequently, its existence and properties are primarily supported by theoretical studies, including molecular dynamics simulations \cite{Lin_2020, li2021tasting, watanabe2007dynamical}, compressible liquid-drop models \cite{Carreau_2019, NEWTON2022137481, Dinh_2021, Parmar_pasta}, Thomas–Fermi approaches \cite{ Furtado_2021, bcpm}, and nuclear density functional theory \cite{schuetrumpf2016clustering}, all of which consistently predict pasta phases near the crust–core transition density. There have also been investigations indicating that nuclear pasta phases are significantly less abundant once shell and pairing effects are properly taken into account \cite{chamel_pasta_shell}.

One of the most promising observational probes of crustal microphysics, and in particular of nuclear pasta, is provided by the quasi-periodic oscillations (QPOs) detected in the X-ray tails of giant flares from soft gamma repeaters (SGRs). These QPOs, observed over a broad frequency range from tens of hertz up to kilohertz, are widely interpreted as manifestations of global oscillation modes of strongly magnetised neutron stars, with the low-frequency components commonly associated with torsional shear oscillations of the crust \cite{Samuelsson2007MNRAS374, Sotani2013MNRAS434, Passamonti2014MNRAS438, SteinerWatts2009PRL103}. The frequency of fundamental torsional oscillations of neutron star crusts has also been used to probe the equation of state \cite{Sotani_2012_prl}. The frequencies and spectral structure of these modes depend sensitively on the crustal shear modulus, effective mass density, entrainment, and crust thickness, all of which can be significantly modified near the crust–core boundary by the emergence of nuclear pasta. In particular, the expected reduction and possible anisotropy of the elastic response in the pasta layer can shift mode frequencies, modify mode spacing, and enhance the coupling between crustal and core magneto-elastic oscillations, thereby affecting both the persistence and observability of QPOs during magnetar flares \cite{Gearheart_2011, Piro_2005, Passamonti2016MNRAS463}. While such effects make QPOs a potentially powerful indirect probe of nuclear pasta, their interpretation is complicated by uncertainties in the equation of state, magnetic-field configuration, superfluid properties, and the extent and mechanical response of the pasta phase itself. This complexity makes it essential to investigate the role of nuclear pasta by explicitly incorporating uncertainties in the equation of state when comparing theoretical predictions with QPO observations.

Only a limited number of studies have investigated the nuclear pasta phase within a statistical framework that explicitly accounts for uncertainties in the underlying nuclear equation of state. Recent Bayesian and statistical attempts to characterize pasta properties were carried out by \cite{NEWTON2022137481} and \cite{Dinh_2021}, which employed Skyrme-based energy density functionals and metamodel descriptions of dense matter within the compressible liquid-drop model (CLDM) to explore the onset and extent of pasta phases.  In this work, we build upon our recent Bayesian analysis of unified neutron-star equations of state, in which the core and crust were consistently described within a relativistic mean-field (RMF) framework and constrained by a $\approx$ 20 combination of nuclear experimental data, empirical saturation properties, chiral effective field theory at low densities, and multimessenger astrophysical observations \cite{Parmar:2026bmm}. We extend the resulting posterior ensemble to investigate the emergence of nuclear pasta phases near the crust–core interface and to quantify the associated theoretical uncertainties. Using these posteriors, we systematically study the properties of the pasta layer, including its onset density, thickness, and mass fraction, and explore their correlations with underlying nuclear-matter parameters such as the symmetry energy and its density dependence. Attempts have also been made to constrain nuclear pasta using quasi-periodic oscillation (QPO) observations of magnetars \cite{Sotani_2012_prl, Sotani2013MNRAS434}. However, such studies typically rely on a small number of representative equations of state or simplified crust models, although these equations of state span a range of nuclear-matter properties. Here, instead, we exploit the full posterior distribution of unified RMF equations of state to systematically propagate current uncertainties in the neutron-star equation of state into predictions for crustal shear properties and torsional oscillation spectra. Following \cite{SteinerWatts2009PRL103}, we compute QPO frequencies and assess the impact of pasta-induced modifications of the crust on these modes while systematically propagating current uncertainties in the neutron-star equation of state.

The  manuscript is organized as follows. We first briefly outline the formalism adopted in this work, including the description of nuclear pasta within the compressible liquid-drop model (CLDM) and the determination of crustal torsional oscillation modes used to compute QPO frequencies following \cite{SteinerWatts2009PRL103}, in Sec. \ref{formalism}. Detailed discussions of the CLDM framework, the treatment of pasta phases, 
can be found in Ref.\cite{NEWTON2022137481,Dinh_2021, Parmar_pasta}. The relativistic mean-field (RMF) models employed, as well as the Bayesian methodology and the constraints adopted in this study motivated by \cite{Tsang2024}, can be found in Refs.\cite{Parmar:2026bmm}. We then present our results in Sec.\ref{results} and summarize our main findings and conclusions in Sec.~\ref{conclusion}.

\section{\label{formalism} Formalism}
\subsection{ Compressible Liquid Drop Model (CLDM) for nuclear pasta}
\label{cldmform}
In CLDM formalism, the energy of the system in the inner crust using Wigner-Seitz (WS) approximation can be written as \cite{Dinh_2021},
\begin{align}
    E(r_c, y_p,n_i, n_n)&=f(u)\left[E_{\rm bulk}(n_i, y_p)\right]
    \nonumber\\
    &
    +E_{\rm bulk}(n_g,0)\left[1-f(u)\right]
    \nonumber\\
    &
    +E_{\rm surf}+E_{\rm curv}+E_{\rm coul}+E_e.
    \label{eq:energy}
\end{align}
Here $r_c$ is the radius (half-width in the case
of planar geometry) of WS cell, $y_p$ the proton fraction and $n_i$ and $n_g$ are the baryon density of charged nuclear component and density of neutron gas respectively.  The function $f(u)$ define the volume fraction occupied by the cluster or hole \cite{Parmar_pasta, Dinh_2021}.  $E_{bulk}$ represents the bulk energy, which in the present work is calculated within the RMF framework.

Pasta structure only affects the finite size effects which can be expressed analytically as a function of the dimension of the  pasta structure. We consider the three canonical geometries namely spherical, cylindrical, and planar, defined by a dimensionality parameter $d = 3, 2, 1,$ respectively. We then define the finite size corrections along the same lines as in \cite{Newton_2013, Dinh_2021}. The surface and curvature energies are written as \cite{Newton_2013, Dinh_2021},
\begin{equation}
    E_{\rm surf}+E_{\rm curv}=\frac{u d}{r_N}\left( \sigma_s +\frac{(d-1)\sigma_c}{r_N}\right),
\end{equation}
where $r_N$ is the radius/half-width of the cluster/hole and   $\sigma_s$ and $\sigma_c$ are the dimension independent surface and curvature tension  based on the Thomas-Fermi calculations \cite{Ravenhall_1983} and re writen as

\begin{equation}
\label{surf}
    \sigma_s=\sigma_0\frac{2^{p+1} + b_s}{y_p^{-p} +b_s+(1-y_p)^{-p}},
\end{equation}
\begin{equation}
\label{curv}
    \sigma_c=\gamma \, \sigma_s\frac{\sigma_{0,c}}{\sigma_0}\left(\beta-y_p\right).
\end{equation}
Here the parameters ($\sigma_0$, $\sigma_c$, $b_s$, $\gamma$, $\beta$, $p$) are optimised for a given equation of state on the atomic mass evaluation 2020 data \cite{Huang_2021}. The Coulomb energy reads as \cite{Dinh_2021}
\begin{equation}
    E_{\rm coul}=2\pi (e\,y_p\,n_i\,r_N)^2\,u\,\eta_d(u),
\end{equation}
where e is the elementary charge and $\eta_d(u)$ is associated with the pasta structure as \cite{Dinh_2021, Newton_2013}
\begin{equation}
    \eta_d(u)=\frac{1}{d+2}\Big[\frac{2}{d-2}\Big(1-\frac{du^{1-\frac{2}{d}}}{2}\Big)+u\Big]
\end{equation}
for $d=1$ and 3 whereas for $d=2$ it reads as,
\begin{equation}
    \eta_d(u)=\frac{1}{4}\Big[\log (\frac{1}{u}) + u -1\Big].
\end{equation}
For a given baryon density, the equilibrium composition of a WS cell is obtained by minimising the energy per unit volume using the variational method where the auxiliary function to be minimized reads as \cite{Parmar_2022, Carreau_2019}
\begin{equation}
    \mathcal{F}=\frac{E_{WS}}{V_{WS}}-\mu_b\rho.
\end{equation}
Here, $\mu_b$ is the baryonic chemical potential. This results in a set of four differential equations   corresponding to mechanical, dynamical, $\beta$-equilibrium and the nuclear virial theorem \cite{Carreau_2020, Carreau_2019}. The viral relation is used to numerically solve the value of $r_N$. To obtain the most stable pasta structure at a given baryon density, we first calculate the composition of a spherical nucleus. Then keeping this composition fixed, we calculate the radius or half-width of five different pasta structure namely, sphere, rod, plate, tube, and bubble. The equilibrium phase is then the one that minimizes the total energy of the system. 
\subsection{Crustal torsional shear-mode frequencies}
\label{sec:qpo}

Neutron stars support a rich spectrum of oscillations governed by different restoring forces, including elastic stresses in the crust, magnetic tension, and fluid pressure gradients. Among these, torsional shear oscillations of the crust have long been recognized as particularly relevant for interpreting the quasi-periodic oscillations (QPOs) observed during magnetar flares \cite{Duncan1998,Samuelsson2007}. These modes are incompressible, require comparatively low excitation energy, and can persist over timescales compatible with the observed QPO durations \cite{Israel2005,Strohmayer2005}. Moreover, crustal motion is naturally expected in flare scenarios in which magnetic stresses progressively load the solid crust until yielding occurs, thereby exciting global oscillations that can couple to the external magnetosphere and modulate the observed X-ray emission \cite{Thompson1996,Link2014}. While strong magnetic fields inevitably couple crustal shear modes to Alfv\'en-like oscillations in the core, detailed magneto-elastic studies have shown that the characteristic frequencies associated with crustal shear motion can still emerge in the coupled system, particularly at low frequencies \cite{Glampedakis2006,Levin2007,Gabler2013}.

Following \cite{SteinerWatts2009PRL103,Piro_2005}, we compute the frequencies of
crustal torsional (shear) oscillations using a plane-parallel approximation, which
provides an accurate description of axial shear modes while substantially
simplifying the eigenvalue problem. In this approach, the crust is treated as a
vertically stratified slab with coordinate $z$, and we consider purely toroidal,
incompressible displacements with no vertical component. Assuming a harmonic time
dependence, $\xi(z,t)=\xi(z)\,e^{i\omega t}$, the perturbation equation for the
azimuthal displacement $\xi(z)$ reduces to the one-dimensional eigenvalue equation
\cite{SteinerWatts2009PRL103}
\begin{equation}
\label{eq:steiner_eq}
\frac{d}{dz}\!\left(\mu\,\frac{d\xi}{dz}\right)
+\rho_d\,\omega^2\!\left(1+\frac{v_A^2}{c^2}\right)\xi
-\mu\,k_\perp^2\,\xi
=0 ,
\end{equation}
where $\mu$ is the shear modulus, $\rho_d$ is the dynamical mass density, $v_A=B^z/\sqrt{4\pi\rho_d}$ is the Alfv\'en speed associated
with a uniform vertical magnetic field $B^z$, and $k_\perp$ denotes the transverse
wavenumber associated with the angular structure of the mode. The plane-parallel geometry neglects curvature effects inherent to spherical stars.
To recover the correct torsional-mode spectrum in the zero-field limit, we follow
\cite{SteinerWatts2009PRL103,Samuelsson2007MNRAS374} and mimic spherical geometry by
identifying the transverse wavenumber as
\begin{equation}
\label{eq:kperp}
k_\perp^2 = \frac{(l+2)(l-1)}{R^2},
\end{equation}
where $l$ is the angular quantum number of the torsional mode and $R$ is the stellar
radius. 

The boundary conditions correspond to vanishing traction at the stellar surface and
the disappearance of shear stresses at the crust--core interface, namely
\begin{equation}
\left.\frac{d\xi}{dz}\right|_{\rm surf}=0,
\qquad
\left.\mu\,\frac{d\xi}{dz}\right|_{\rm cc}=0.
\end{equation}
In the present work, we restrict our analysis to the non-magnetic case ($B^z=0$),
for which Eq.~(\ref{eq:steiner_eq}) reduces to a purely elastic eigenvalue problem
that determines the crustal torsional shear-mode frequencies.

The shear modulus of a body-centered cubic Coulomb lattice embedded in a uniform electron background, in the low-temperature limit and including electron-screening corrections inferred from Monte Carlo simulations \cite{Chugnov_2010}, is given by \cite{tews_2017, Sotani_2013}
\begin{equation}
\label{eq:shearmodulus}
    \mu = 0.1194\left(1-0.010\,Z^{2/3}\right)
    \frac{n_i\left(Ze\right)^2}{a},
\end{equation}
where $n_i$ is the ion mass density, $Ze$ is the nuclear charge, and $a=R_{\rm WS}$ is the Wigner--Seitz radius. Equation~(\ref{eq:shearmodulus}) is valid for spherical nuclei. However, near the crust--core boundary, non-spherical nuclear configurations (nuclear pasta) may appear, for which the elastic properties remain uncertain. It is generally expected that the rigidity of matter decreases in this region \cite{Newton_2022}.

To model the softening of the crust in the pasta layer, extending over the density range $\rho_{ph}\le\rho_b\le\rho_c$, where $\rho_{ph}$ denotes the onset density of non-spherical shapes and $\rho_c$ the crust--core transition density, we adopt a phenomenological prescription in which the effective shear modulus is smoothly suppressed according to \cite{Sotani_2012, Gearheart_2011, Passamonti_2016}
\begin{equation}
\label{eq:mubar}
    \bar{\mu} = c_1\left(\rho_b-\rho_c\right)^2\left(\rho_b-c_2\right),
\end{equation}
where the constants $c_1$ and $c_2$ are fixed by requiring $\bar{\mu}$ to match Eq.~(\ref{eq:shearmodulus}) continuously at $\rho_b=\rho_{ph}$ and to vanish smoothly at $\rho_b=\rho_c$. The latter condition ensures that the shear speed approaches zero at the crust--core interface. The local shear speed is then defined as  \cite{tews_2017}
\begin{equation}
\label{eq:shearspeed}
    v_s = \sqrt{\frac{\mu}{\rho_d}},
\end{equation}
 Neglecting the effects of neutron superfluidity and entrainment, we set $\rho_d$ equal to the total mass density, $\rho_d=\rho_m$ \cite{SteinerWatts2009PRL103}. This prescription allows us to consistently compute torsional mode frequencies while isolating the impact of pasta-induced softening of the crust.

\begin{figure*}
    \centering
    \includegraphics[width=1\linewidth]{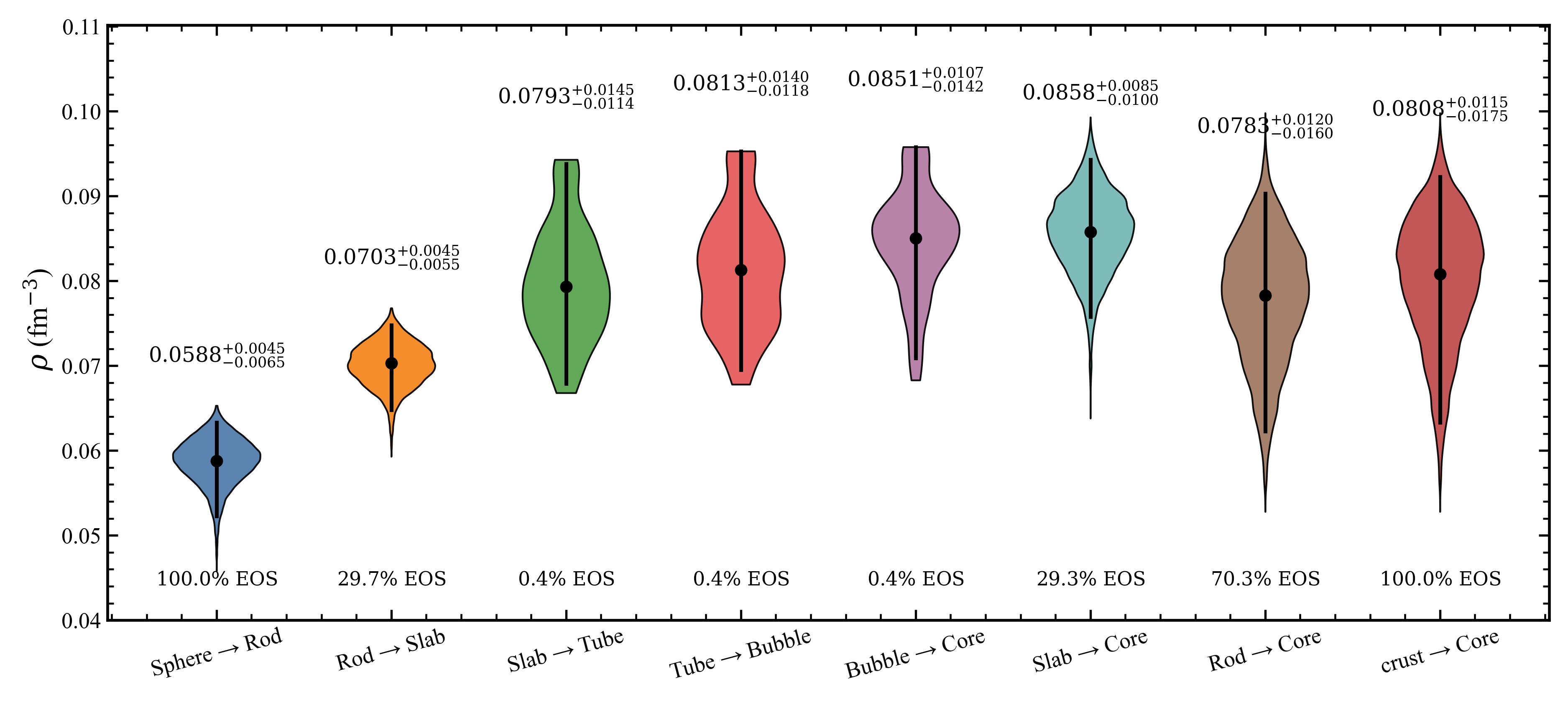}
    \caption{
Posterior distributions of the transition densities between different nuclear pasta geometries obtained within the CLDM framework using RMF parameter sets drawn from the Bayesian posterior.
The violin plots summarize the probability distributions of the transition densities between successive geometries, with the width indicating the relative posterior weight.
}
    \label{fig:pasta_posterior}
\end{figure*}

\section{\label{results} Results}
Using the posterior distribution of 
RMF parameter sets constrained by nuclear experimental data, chiral effective field theory at sub-saturation densities, and multimessenger neutron-star observations \cite{Parmar:2026bmm}, we compute the nuclear pasta structure within the compressible liquid-drop model (CLDM). In this approach, we assume a fixed set of five  geometries, spherical nuclei, cylindrical rods, planar slabs, cylindrical tubes, and spherical bubbles, which are commonly adopted in the literature and are motivated by molecular-dynamics simulations of dense nuclear matter \cite{Oyamatsu_2007, Caplan_2017}. For each RMF parameter set in the posterior ($\approx 40000$), the bulk nuclear and neutron-matter properties entering the CLDM are provided by the RMF equation of state. At each baryon density, the total energy density is evaluated for all five geometries, including bulk, surface, curvature, and Coulomb contributions, and the energetically favoured configuration is selected by minimizing the total energy. This procedure is repeated over the full density range of the inner crust, yielding a consistent determination of the pasta sequence and transition densities for each RMF parameter set. We fit the surface and curvature terms for each RMF equation of state,  on AME2020 mass table \cite{Huang_2021}. The details regarding this fitting is given in \cite{Parmar:2026bmm}.

\begin{figure*}
    \centering
    \includegraphics[width=1\linewidth]{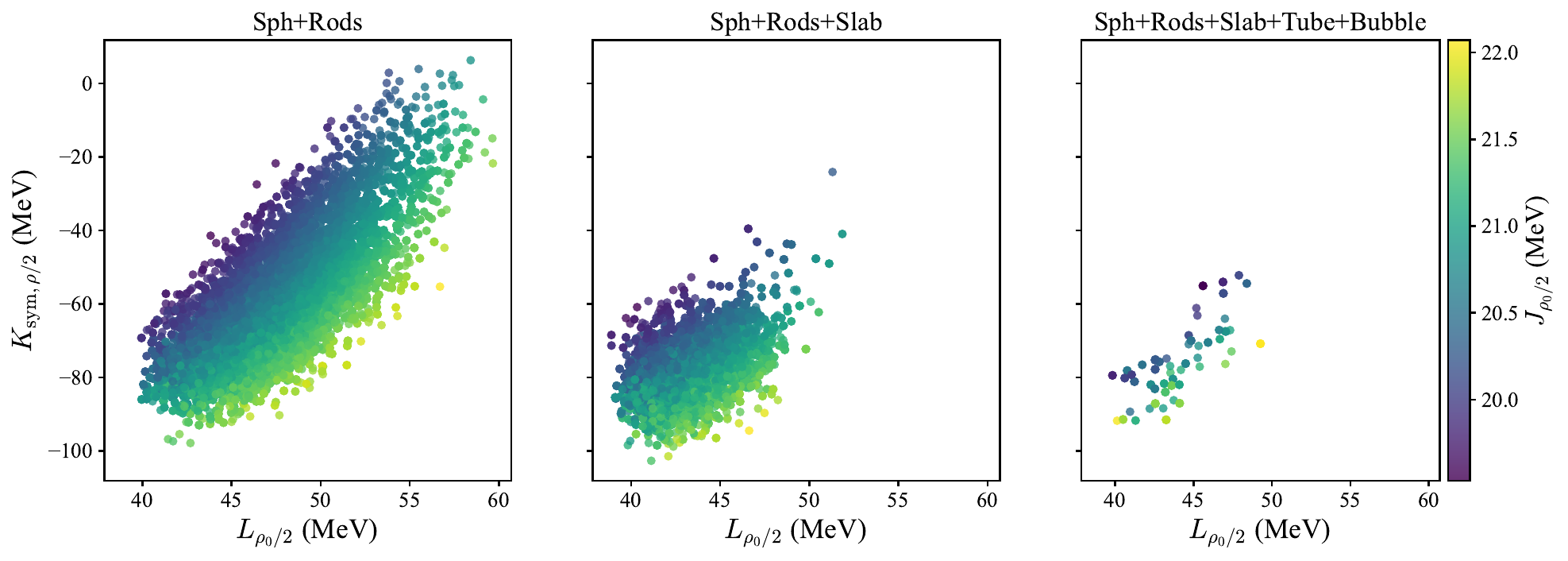}
    \caption{
Correlation between the appearance of nuclear pasta geometries and the symmetry-energy
parameters evaluated at half the saturation density, $L(\rho_0/2)$,
$K_{\mathrm{sym}}(\rho_0/2)$, and $J(\rho_0/2)$. The equations of state are classified into
three categories, guided by Fig.~\ref{fig:pasta_posterior}: models exhibiting only
spherical nuclei and cylindrical rods; models that additionally favor the slab geometry; and models that predict further pasta
geometries beyond slabs, including tubes and bubbles.}
\label{fig:diff_pasta}
\end{figure*}

Figure~\ref{fig:pasta_posterior} summarizes the posterior distributions of the transition
densities associated with the formation and evolution of nuclear pasta structures in the
inner crust. We show the distributions of the densities at which the system undergoes
transitions between the assumed CLDM geometries, with the energetically favored
configuration at each density. We also present the number of equations of state that support a given pasta geometry, together with the corresponding 95\% credible range for the density at which each pasta configuration first appears. The first transition,
from spherical nuclei to cylindrical rods, is tightly constrained with 
$\rho_{sr} \simeq 0.0588^{+0.0045}_{-0.0065} ~\mathrm{fm}^{-3}$. This transition is of particular importance, as it marks
the onset of the pasta layer, which is expected to exhibit a disordered or glassy character
and a significantly reduced effective rigidity compared to the crystalline lattice of
spherical nuclei \cite{Newton_2022}. As a result, the density and pressure at this transition play a key role in
determining the global properties of the pasta region. The subsequent evolution of the pasta structure depends sensitively on the density
dependence of the symmetry energy, $J(\rho)$ as is shown on \cite{Parmar_pasta, Oyamatsu_2007}. It was shown in \cite{Dinh_2021} that surface and curvature terms play a significant role in determining nuclear pasta properties, and that using a surface tension inconsistent with the bulk energy leads to an underestimation of the pasta phase. In the present work, we address this issue by consistently fitting the bulk energy together with the surface and curvature terms to atomic mass table data, as described earlier. We therefore expect the resulting description to provide a self-consistent and reliable estimate of pasta properties. After the rod phase, the majority of the
posterior samples ($\sim 70\%$) undergo a direct transition from rods to uniform core
matter, with no intermediate geometries being energetically favored. Nevertheless, all
parameter sets predict at least one non-spherical phase beyond spherical nuclei, i.e. rods. For
approximately $30\%$ of the equations of state, the system proceeds to the slab geometry
following the rod phase. Direct transitions beyond the slab phase are rare: only about
$0.4\%$ of the posterior samples favor a further transition to cylindrical tubes, while the
probability of reaching the bubble phase before the crust--core transition is found to be
negligible.

The appearance and extent of pasta layers have a direct impact on the density at which the
crust--core transition occurs. In Ref.~\cite{Parmar:2026bmm}, we found the crust--core
transition density to be
$\rho_t = 0.0788^{+0.0066}_{-0.0086}~\mathrm{fm}^{-3}$, considering spherical geometry. When the transition from rods directly
to the core is considered, this value remains largely unchanged. In contrast, for equations
of state that predict additional pasta geometries, the crust--core transition density shifts
to slightly higher values, reflecting the lower energy of non-spherical configurations that
delay the onset of uniform matter. Even in these cases, however, the transition density never
exceeds $\rho_t \simeq 0.1~\mathrm{fm}^{-3}$. Overall, the transition densities from spherical
nuclei to rods and from rods to slabs are particularly well constrained, with posterior peaks
at $\rho \simeq 0.0588~\mathrm{fm}^{-3}$ and $\rho \simeq 0.0703~\mathrm{fm}^{-3}$, respectively.

It has been widely argued in the literature that the properties of the neutron-star crust
are predominantly controlled by the isovector sector of the nuclear equation of state,
in particular by the slope and curvature of the symmetry energy, commonly quantified
by the parameters $L$ and $K_{\mathrm{sym}}$ \cite{universe7060182, Chamel_2012, Pearson_2018, parmar2021crustal, Dinh_2021}. In this context, the appearance and evolution
of nuclear pasta geometries are also expected to be regulated primarily by these
quantities. Furthermore, it has been suggested that nuclear-matter properties at
sub-saturation densities play a particularly important role in determining crustal
structure \cite{Parmar:2026bmm, Balliet_2021, Ducoin_2011}. Motivated by these considerations, we investigate the connection between
nuclear-matter parameters and the emergence of pasta phases, with the aim of
identifying which properties most strongly influence the pasta structure.

Figure~\ref{fig:diff_pasta} illustrates the correlation between the appearance of pasta
geometries and the symmetry-energy parameters evaluated at half the saturation density,
namely $L(\rho_0/2)$, $K_{\mathrm{sym}}(\rho_0/2)$, and $J(\rho_0/2)$. We use $(\rho_0/2)$ to calculate the nuclear matter properties, as it shows the highest correlation with the  crust properties \cite{Parmar:2026bmm}, and also the applicable density range are in the subsaturation region. Guided by the results
presented in Fig.~\ref{fig:pasta_posterior}, we classify the equations of state into three
distinct categories. The first category consists of equations of state that exhibit only
spherical nuclei and cylindrical rods, a feature shared by all models considered. The
second category includes those equations of state that, in addition to spheres and rods,
also favor the slab geometry, which occurs for approximately $30\%$ of the posterior
samples. The third category comprises equations of state that predict the appearance of
additional pasta geometries beyond slabs, including tubes and bubbles.

It is evident from Fig.~\ref{fig:diff_pasta} that while the first category, containing only
spherical nuclei and rods, spans a broad range of values in the $L$--$K_{\mathrm{sym}}$
plane, the appearance of the slab phase is strongly associated with relatively low values
of $L(\rho_0/2)$ and  large negative values of
$K_{\mathrm{sym}}(\rho_0/2)$. The same region of parameter space also hosts the small
subset of equations of state that predict the full sequence of pasta geometries. For the
sphere--rod category, we further observe that lower values of the symmetry energy
$J(\rho_0/2)$ tend to correlate with less negative values of
$K_{\mathrm{sym}}(\rho_0/2)$ at a given $L(\rho_0/2)$. To further quantify the relative influence of nuclear-matter parameters on the appearance
of pasta structures, we employ a supervised machine-learning approach based on a
decision tree classifier, implemented using \texttt{sklearn.tree.DecisionTreeClassifier}. The classifier identifies the nuclear-matter parameters that most effectively discriminate
between different pasta geometries by recursively partitioning the parameter space \cite{Breiman1984CART}.
Using $L(\rho_0/2)$, $K_{\mathrm{sym}}(\rho_0/2)$, and $J(\rho_0/2)$ as input features, we
obtain the following feature importances:
\[
\left[ L (\rho_0/2),\, K_{\mathrm{sym} }(\rho_0/2),\, J(\rho_0/2) \right]
=
\left[ 0.77,\; 0.22,\; 0.01 \right].
\]
This result indicates that the slope of the symmetry energy at sub-saturation densities is
the dominant factor controlling the emergence of pasta phases, with the curvature playing
a secondary but non-negligible role, while the absolute magnitude of the symmetry energy
has only a marginal impact.

{
\renewcommand{\arraystretch}{1.4}

\begin{table}[t]

\caption{Median values and 95\% credible intervals (CI) for pasta-related quantities obtained in the present Bayesian analysis. The second and third columns list, where available, corresponding results from the PNM+Skins analysis of Newton \textit{et al.}~\cite{NEWTON2022137481} and from Dinh Thi \textit{et al.}~\cite{Dinh_2021}, respectively. Here, $\rho_{\rm sr}$, $P_{\rm sr}$, and $\mu_{\rm sr}$ denote the baryon density, pressure, and chemical potential at the spherical-to-rod transition, while $\alpha_{\rm sr}$ and $\alpha_{\rm core}$ represent the isospin asymmetry ($\alpha= \frac{\rho_n- \rho_p}{\rho_n+ \rho_p}$) at the spherical-to-rod transition and at the crust--core boundary. The quantities $\rho_t$, $P_t$, and $\mu_t$ correspond to the density, pressure, and chemical potential at the crust--core transition. The ratios $\Delta R_{\rm pasta}/\Delta R_{\rm c}$ and $\Delta M_{\rm pasta}/\Delta M_{\rm c}$ give the fractional thickness and mass of the entire pasta layer relative to the crust, whereas $\Delta R_{\rm rod}/\Delta R_{\rm c}$ and $\Delta M_{\rm rod}/\Delta M_{\rm c}$ refer specifically to the rod phase.}
\label{tab:pasta_ci}
\begin{ruledtabular}

\begin{tabular}{lccc}
Quantity
& This work
& \cite{NEWTON2022137481}
& \cite{Dinh_2021} \\

\hline
$\rho_{\rm sr}\;[\mathrm{fm}^{-3}]$
& $0.0588^{+0.0045}_{-0.0065}$
& 
&  \\

$P_{\rm sr}\;[\mathrm{MeV\,fm^{-3}}]$
& $0.252^{+0.055}_{-0.055}$
& 
&  \\

$\mu_{\rm sr}\;[\mathrm{MeV}]$
& $11.04^{+1.54}_{-1.89}$
& 
&  \\

$\alpha_{\rm sr}$
& $0.795^{+0.034}_{-0.037}$
& $0.770^{+0.034}_{-0.032}$
&  \\

$\alpha_{\rm core}$
& $0.923^{+0.014}_{-0.014}$
& $0.918^{+0.012}_{-0.014}$
&  \\

$\rho_{t}\;[\mathrm{fm}^{-3}]$
& $0.0808^{+0.0115}_{-0.0175}$
& 
&  \\

$P_{t}\;[\mathrm{MeV\,fm^{-3}}]$
& $0.481^{+0.118}_{-0.143}$
& $0.33^{+0.07}_{-0.07}$
&  \\

$\mu_{t}\;[\mathrm{MeV}]$
& $14.28^{+2.31}_{-2.78}$
& $12.6^{+1.8}_{-1.9}$
&  \\

$\Delta R_{\rm pasta}/\Delta R_{\rm c}$
& $0.140^{+0.025}_{-0.036}$
& $0.129^{+0.019}_{-0.030}$
& $0.128^{+0.047}_{-0.047}$ \\

$\Delta M_{\rm pasta}/\Delta M_{\rm c}$
& $0.475^{+0.071}_{-0.113}$
& $0.54^{+0.05}_{-0.09}$
& $0.485^{+0.138}_{-0.138}$ \\

$\Delta R_{\rm rod}/\Delta R_{\rm c}$
& $0.085^{+0.023}_{-0.031}$
& 
&  \\

$\Delta M_{\rm rod}/\Delta M_{\rm c}$
& $0.307^{+0.073}_{-0.109}$
& 
&  \\
\end{tabular}
\end{ruledtabular}
\end{table}
}

Having identified the presence and sequence of pasta layers, it is essential to assess
their relative contribution to the global crustal properties. In particular, the thickness
and mass fraction of the pasta region directly affect crustal rigidity, transport
properties, and the effective moment of inertia. Previous studies have shown that the
relative thickness and mass of the neutron-star crust are correlated with  the
stellar mass, radius, and a single parameter associated with the crust--core interface
\cite{Lattimer_2007,Zdunik_2017}. Along similar lines, Newton \textit{et al.}
\cite{NEWTON2022137481} used simple expressions for the relative thickness and mass of a
pasta layer, given by
\begin{equation}
\label{eq:rr}
    \frac{\Delta R_p}{\Delta R_c} \approx
    \frac{\mu_c - \mu_p}{\mu_c - \mu_0},
\end{equation}
\begin{equation}
\label{eq:pp}
    \frac{\Delta M_p}{\Delta M_c} \approx
    1 - \frac{P_p}{P_c},
\end{equation}
where $\mu_c$, $\mu_p$, and $\mu_0$ denote the baryon chemical potential at the
crust--core transition, at the onset of the pasta layer, and at the stellar surface,
respectively. The quantities $P_p$ and $P_c$ correspond to the pressure at the bottom of
the pasta layer and at the crust--core transition. Furthermore, since the crustal moment
of inertia is proportional to the crustal mass to first order \cite{Lorenz_1993}, the
relative pasta contribution to the moment of inertia can be approximated as
\begin{equation}
    \frac{\Delta I_p}{\Delta I_c} \approx
    \frac{\Delta M_p}{\Delta M_c}.
\end{equation}

In Table~\ref{tab:pasta_ci}, we report the posterior credible intervals of all relevant
quantities, including the densities, pressures, and chemical potentials at the onset of
the pasta phase and at the crust--core transition, as well as the isospin asymmetry of the
clusters. Using these inputs, we compute the relative pasta thickness
$\Delta R_{\rm pasta}/\Delta R_{\rm c}$ and mass fraction
$\Delta M_{\rm pasta}/\Delta M_{\rm c}$. Our results are compared with those obtained in
Ref.~\cite{NEWTON2022137481}, where these quantities were evaluated within a compressible
liquid-drop model based on an extended Skyrme energy-density functional, constrained by
pure neutron matter calculations, neutron-skin data of ${}^{208}$Pb, and chiral effective
field theory predictions at sub-saturation densities.

It should be noted that in Ref.~\cite{NEWTON2022137481} the surface and curvature
parameters entering the compressible liquid-drop model were sampled using uniform
priors. In contrast, in the present study these parameters are constrained by fitting to
the AME2020 atomic mass evaluation, with the surface parameter fixed to $p=3$ \cite{Dinh_2021, NEWTON2022137481, Parmar:2026bmm}. This
choice has been shown to yield CLDM results that are effectively equivalent to those
obtained within Thomas--Fermi approaches. Furthermore, the present analysis is based on the posterior distribution of relativistic
mean-field (RMF) parameters obtained by fitting to a broad and up-to-date set of
constraints, including nuclear experimental data, empirical saturation properties,
chiral effective field theory predictions at sub-saturation densities, and multimessenger
neutron-star observations. 

From Table~\ref{tab:pasta_ci}, one can see that the isospin asymmetry at the onset of the
pasta phase as well as at the crust--core transition, is consistent with the results of
Ref.~\cite{NEWTON2022137481} within uncertainties. The crust--core transition pressure
$P_t$ obtained in the present work is slightly higher at the level of the posterior median
than that reported by Ref.~\cite{NEWTON2022137481}. 
This difference can be attributed to the requirement of a slightly stiffer equation of state in order to accommodate the NICER mass–radius measurements, as incorporated in the Bayesian estimation of the posterior distributions employed in the present work. While our median estimate predicts marginally thicker pasta layers and a
slightly smaller pasta mass fraction within the crust compared to
Ref.~\cite{NEWTON2022137481}, the two studies remain consistent once the associated
credible intervals are taken into account. In a related Bayesian study of uncertainties in nuclear pasta properties,
Dinh Thi \textit{et al.}~\cite{Dinh_2021} employed a meta-modelling approach to describe
the nuclear equation of state. While the median value of the pasta mass fraction obtained
in the present work is consistent with their results within uncertainties, we find
systematically thicker pasta layers at the level of the posterior median. In Table~\ref{tab:pasta_ci}, we also report the relative thickness and mass of the rod phase of nuclear pasta. This phase is common to all equations of state considered in this study and is seen to constitute the dominant contribution to the overall pasta layer.

\begin{figure*}
    \centering
    \includegraphics[width=0.8\linewidth]{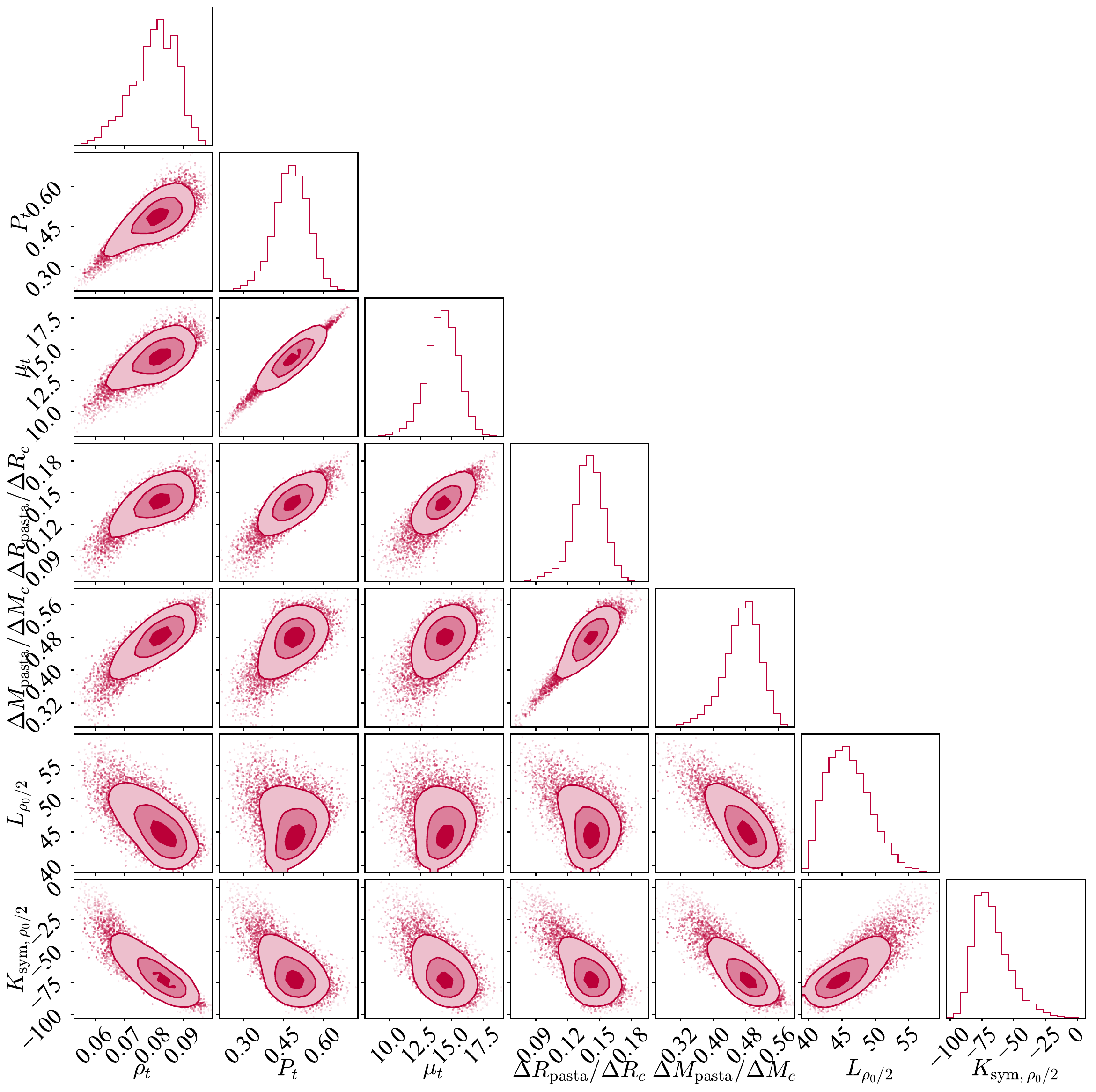}
    \caption{
Corner plot showing the joint posterior distributions of the crust--core transition
pressure and chemical potential, $P_t$ and $\mu_t$, the relative pasta thickness and
mass fractions, $\Delta R_{\rm pasta}/\Delta R_c$ and $\Delta M_{\rm pasta}/\Delta M_c$,
and the symmetry-energy slope and curvature evaluated at sub-saturation density,
$L(\rho_0/2)$ and $K_{\rm sym}(\rho_0/2)$.  
}
\label{fig:corner_pasta_core_LKsym}
\end{figure*}

\begin{figure*}
    \centering
    \includegraphics[width=0.9\linewidth]{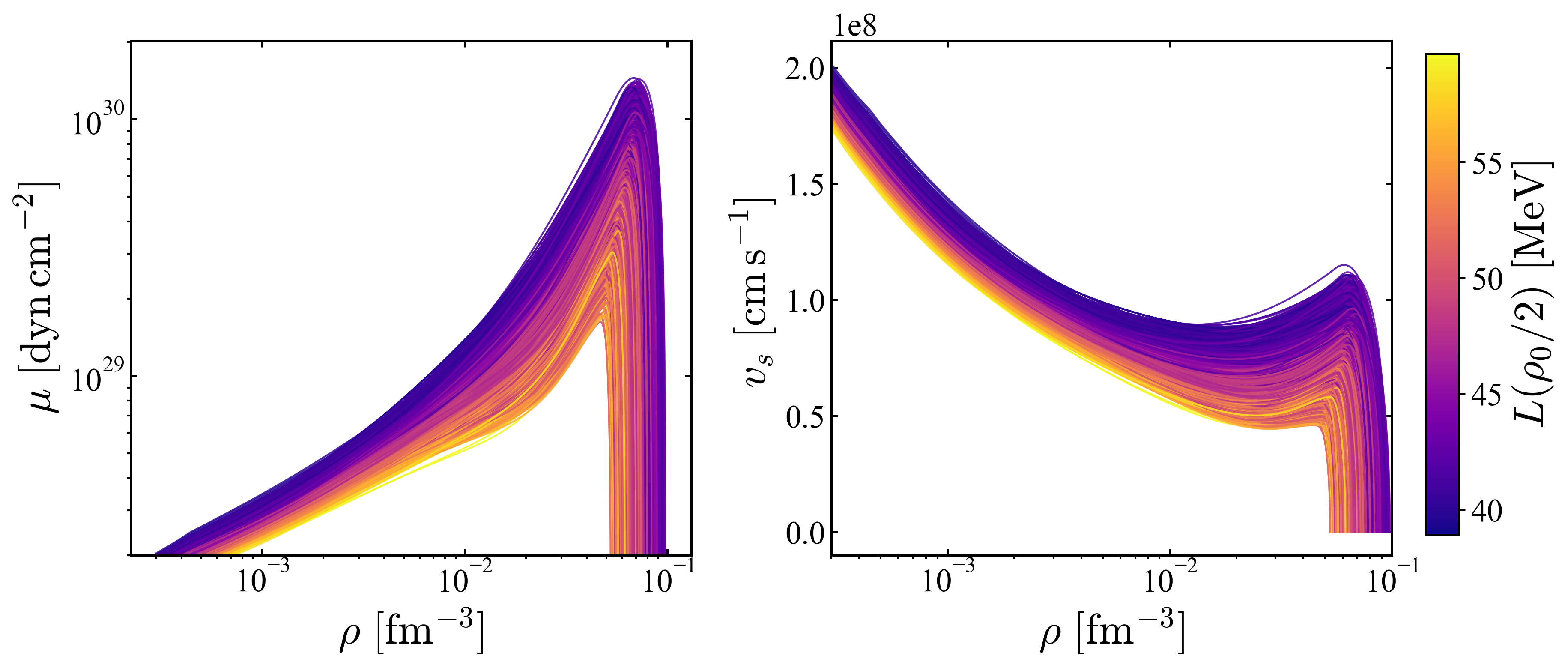}
    \caption{
Posterior distributions of the shear modulus $\mu$ and the corresponding shear-wave
speed $v_s$ throughout the neutron-star crust. For each equation of state drawn from the
Bayesian posterior, the shear modulus is computed using
Eq.~(\ref{eq:shearmodulus}) for the spherical nuclear lattice and is subsequently modified
according to Eq.~(\ref{eq:mubar}) to account for the progressive reduction of rigidity in
the pasta phases. The color coding indicates the value of the symmetry-energy slope
parameter $L$ evaluated at half the saturation density, $L(\rho_0/2)$.
}
\label{fig:shear_posterior}
\end{figure*}

\begin{figure*}
    \centering
    \includegraphics[width=0.9\linewidth]{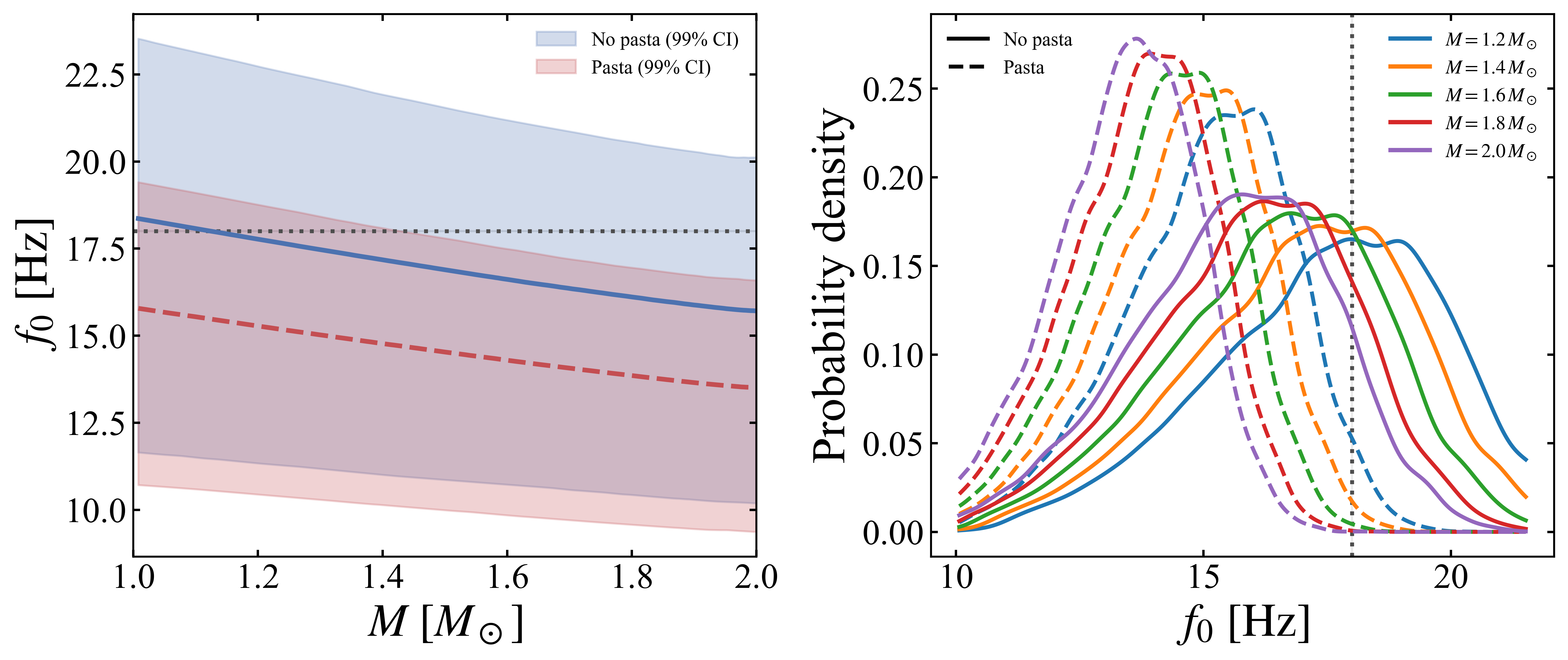}
    \caption{
Left panel: Fundamental crustal torsional oscillation frequencies ($n=0$, $\ell=2$)
as a function of neutron-star mass, shown for models with and without nuclear pasta.
Right panel: Posterior probability distributions of the fundamental crustal frequency
for different neutron-star masses, comparing cases with nuclear pasta included and
neglected. The black dotted line indicates the lowest-frequency QPO observed in
SGR~1806$-$20. The blue and red lines represent the average QPO frequencies for the cases without pasta and with pasta, respectively. }
\label{fig:qpo_fundamental}
\end{figure*}

To assess the dependence of nuclear pasta properties on the underlying nuclear-matter parameters, we examine their correlations with the symmetry-energy slope and curvature. Since it is well established that the symmetry energy at saturation density plays only a limited role in determining the appearance of different pasta geometries as well as the crust–core transition density, we focus instead on the symmetry-energy slope and curvature evaluated at sub-saturation density, $L(\rho_0/2)$ and $K_{\rm sym}(\rho_0/2)$. These quantities more directly control the behavior of matter in the density regime relevant for the formation of nuclear pasta and the crust–core transition. Accordingly, Fig.~\ref{fig:corner_pasta_core_LKsym} presents the joint posterior distributions of the crust–core transition pressure and chemical potential, the relative pasta thickness and mass fractions, and their correlations with $L(\rho_0/2)$ and $K_{\rm sym}(\rho_0/2)$. It can be seen that the crust--core
transition density is strongly correlated with both the relative pasta mass and
thickness, despite the fact that these quantities are not imposed explicitly in the
calculation but are derived from the transition pressure and chemical potential.
This behavior is expected, as the relative pasta mass and thickness are directly
controlled by $P_t$ and $\mu_t$, which determine the depth and extent of the pasta
layer within the crust. Furthermore, the pasta mass and thickness exhibit a noticeably stronger dependence
on the curvature of the symmetry energy, $K_{\rm sym}$, than on the slope parameter
$L$, particularly when evaluated at sub-saturation densities. This is consistent with
previous Bayesian studies, which have shown that $K_{\rm sym}$ plays a dominant role
in shaping the density dependence of the equation of state in the inner crust, while
$L$ primarily influences global crustal properties. The strong correlations with
$P_t$ and $\mu_t$ arise naturally from the thermodynamic relations used to compute
the pasta observables, mirroring trends reported in earlier CLDM-based analyses.

We now turn to the observational implications of nuclear pasta in the inner crust of
neutron stars. In particular, crustal quasiperiodic oscillations (QPOs) are sensitive to
the elastic properties of the crust, most notably to the shear modulus and the
corresponding shear-wave speed. Figure~\ref{fig:shear_posterior} shows the posterior
distributions of the shear modulus and shear speed for all equations of state considered
in this study. Up to the regime of spherical nuclear clusters, the shear modulus can be reliably
approximated using the Monte Carlo–based expression for a body-centered cubic Coulomb
lattice. In the pasta region, where the crystalline order is progressively disrupted, we
model the reduction of rigidity by adopting the effective shear prescription described in
Eq. \eqref{eq:mubar}, which smoothly suppresses the shear modulus as the system
approaches the crust--core transition. We see that the uncertainties in the equation of
state translate into variations of up to a factor of 10 in the shear modulus
throughout the inner crust and into differences of up to a factor of $\sim 2$ in the
maximum shear-wave speed, which is attained near the onset of the rod-like pasta phase. Furthermore, both the shear modulus and the shear-wave speed exhibit a strong
dependence on the symmetry-energy slope parameter evaluated at sub-saturation density,
$L(\rho_0/2)$. Equations of state characterized by smaller values of $L(\rho_0/2)$
generally predict larger shear moduli and higher shear-wave speeds, whereas larger
values of $L(\rho_0/2)$ lead to a softer crust with reduced rigidity and lower shear
speeds.  This was also discussed in \cite{SteinerWatts2009PRL103}, and our large posterior has the same behaviour.

In Fig.~\ref{fig:qpo_fundamental} we illustrate the impact of nuclear pasta on the
fundamental crustal torsional oscillation frequencies. It is evident that including pasta
structures in the inner crust systematically reduces the oscillation frequencies, in
agreement with earlier findings \cite{Gearheart2011MNRAS418}.  In the literature, several attempts have been made to identify the observed crustal
oscillation frequencies in magnetars. The most prominent QPOs detected in
SGR~1806$-$20 occur at 18, 26, 30, and 92.5~Hz, while those observed in
SGR~1900$+$14 appear at 28, 54, and 84~Hz \cite{Watts:2006mr}. Frequencies
above $\sim 100$~Hz have been interpreted not only in terms of higher--multipole
fundamental and overtone shear modes of the crust, but also as polar-type oscillations  .
It has also been suggested that, using  a limited set of equations of state,  the lowest
frequency observed in SGR~1806$-$20 (18~Hz) could be identified with the $\ell=3$
fundamental torsional mode \cite{Sotani_2011} rather than with $\ell=2$.

In the present work, based on a large ensemble of approximately $4\times10^4$ equations
of state constrained by nuclear, theoretical, and astrophysical data \cite{Parmar:2026bmm}, we find that the
$\ell=2$ fundamental crustal mode cannot account for the observed 18~Hz QPO, in agreement with the analysis of \cite{Sotani_2011}  which used arbitary pasta onset densities. This
conclusion is robust across the posterior and implies that, within a purely crustal
framework without crust--core coupling, the 18~Hz oscillation must be associated with
higher--$\ell$ modes if it is to be explained by shear oscillations alone. The right panel of Fig.~\ref{fig:qpo_fundamental} shows the posterior distributions of the
fundamental crustal frequency for several neutron-star masses. When nuclear pasta is
neglected, and a fully spherical-nuclei crust is assumed, only relatively low-mass neutron stars
($M \simeq 1.2$--$1.4\,M_\odot$) marginally allows the 18~Hz frequency to be reproduced.
However, once nuclear pasta is consistently included, the fundamental crustal mode
frequency is shifted to lower values, rendering a purely crustal explanation of the
18~Hz QPO untenable for all masses considered.  In Table~\ref{tab:f0_pasta_ci}, we report the posterior median values and $95\%$ credible
intervals of the fundamental crustal torsional mode frequency $f_0$ for different
neutron-star masses, obtained from our Bayesian ensemble of equations of state with
nuclear pasta included.

\begin{figure}
    \centering
    \includegraphics[width=1\linewidth]{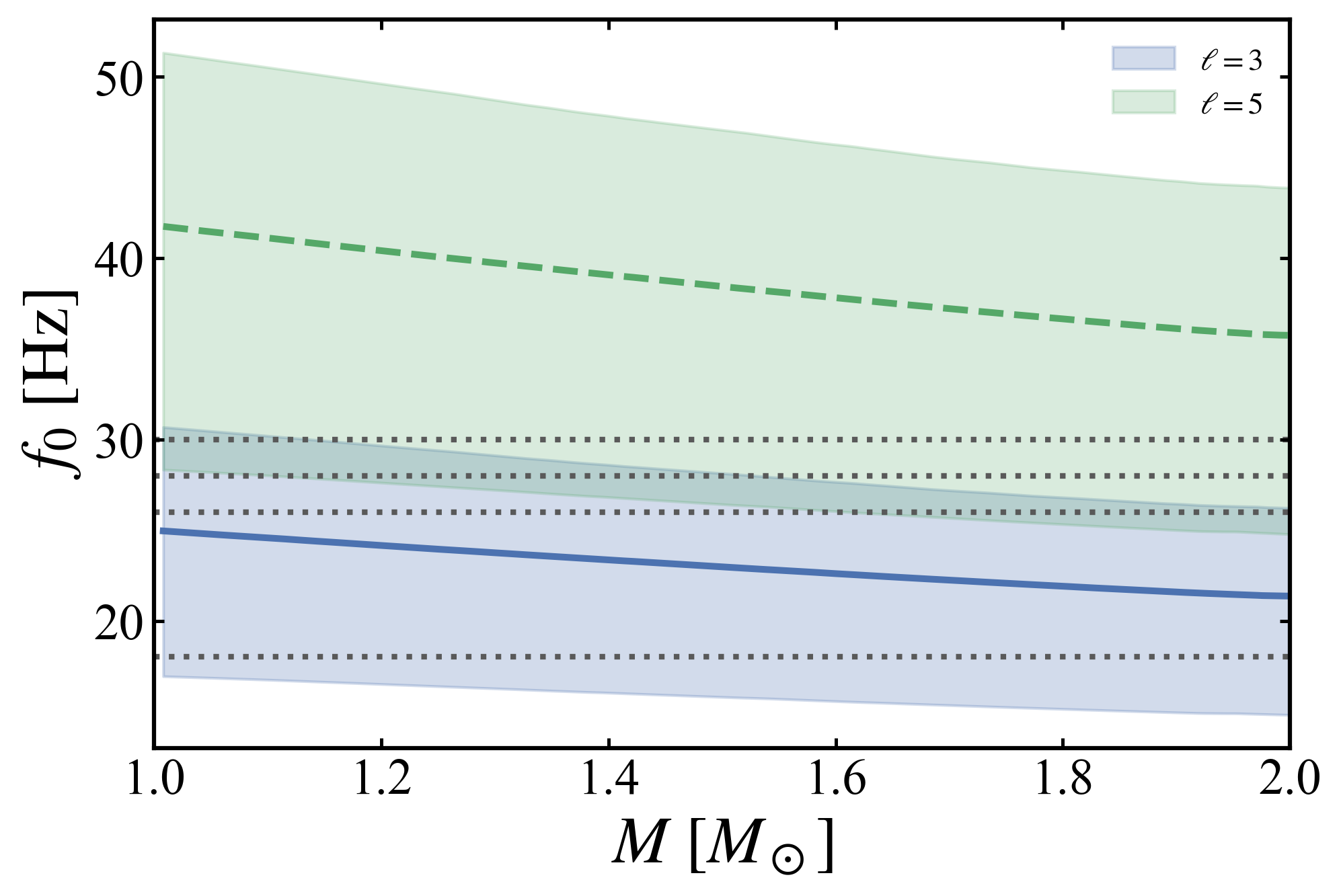}
    \caption{
Posterior distributions of the fundamental crustal torsional mode frequencies for
$\ell=3$ and $\ell=5$ as a function of neutron-star mass. The black dotted horizontal
lines indicate the observed low-frequency QPOs at 18.0, 26.0, 28.0, and 30.0~Hz,
highlighting the mass ranges for which these frequencies can be accommodated within
a purely crustal interpretation.
}
\label{fig:qpo_l3_l5}
\end{figure}
\begin{figure}
    \centering
    \includegraphics[width=1\linewidth]{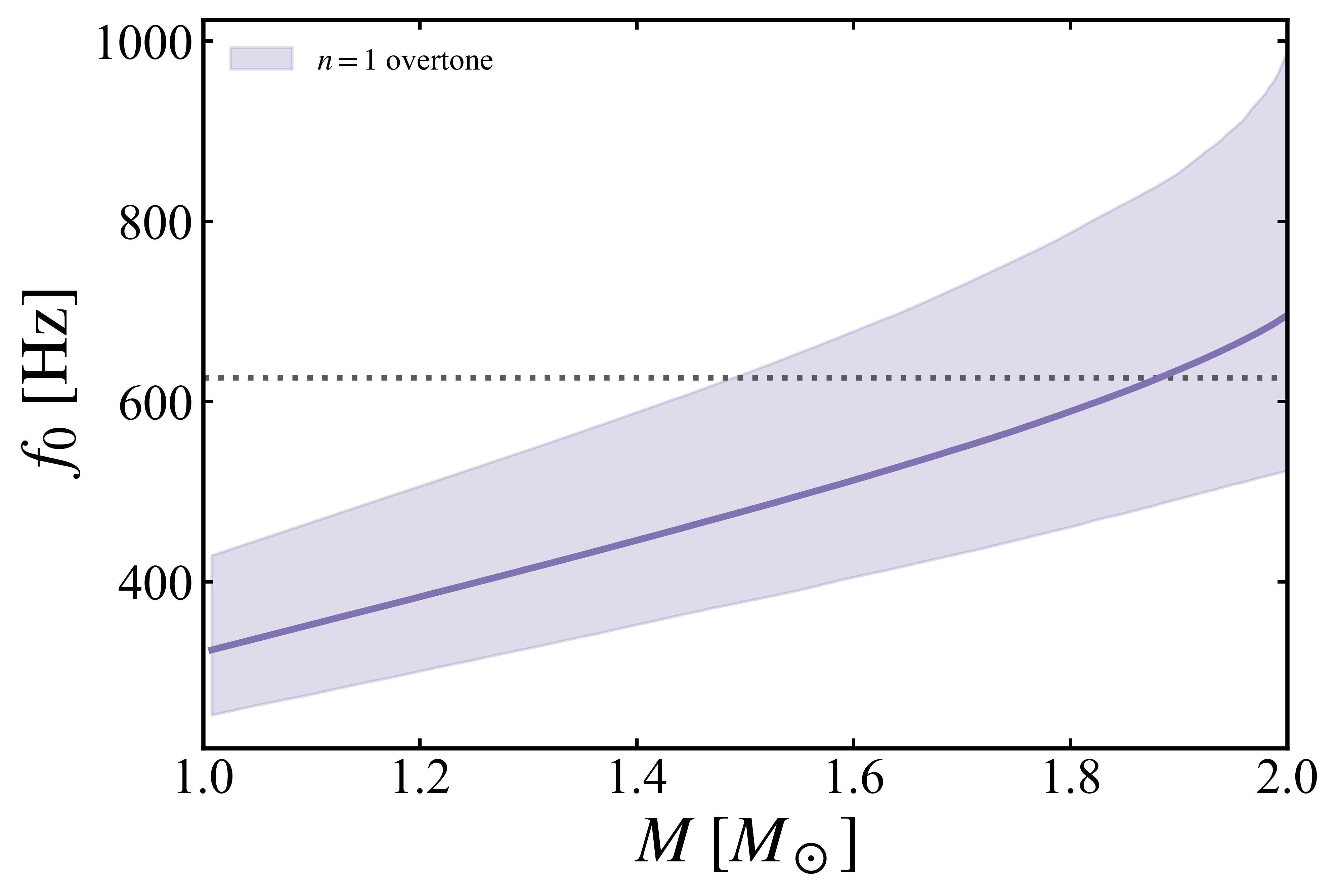}
    \caption{
Posterior distributions of the first overtone crustal torsional mode frequency as a
function of neutron-star mass. The black dotted horizontal line marks the observed
626.5~Hz QPO in SGR~1806$-$20, commonly identified as the first overtone of the crustal
shear oscillation.
}
\label{fig:qpo_overtone}
\end{figure}

\begin{figure*}
    \centering
    \includegraphics[width=0.9\linewidth]{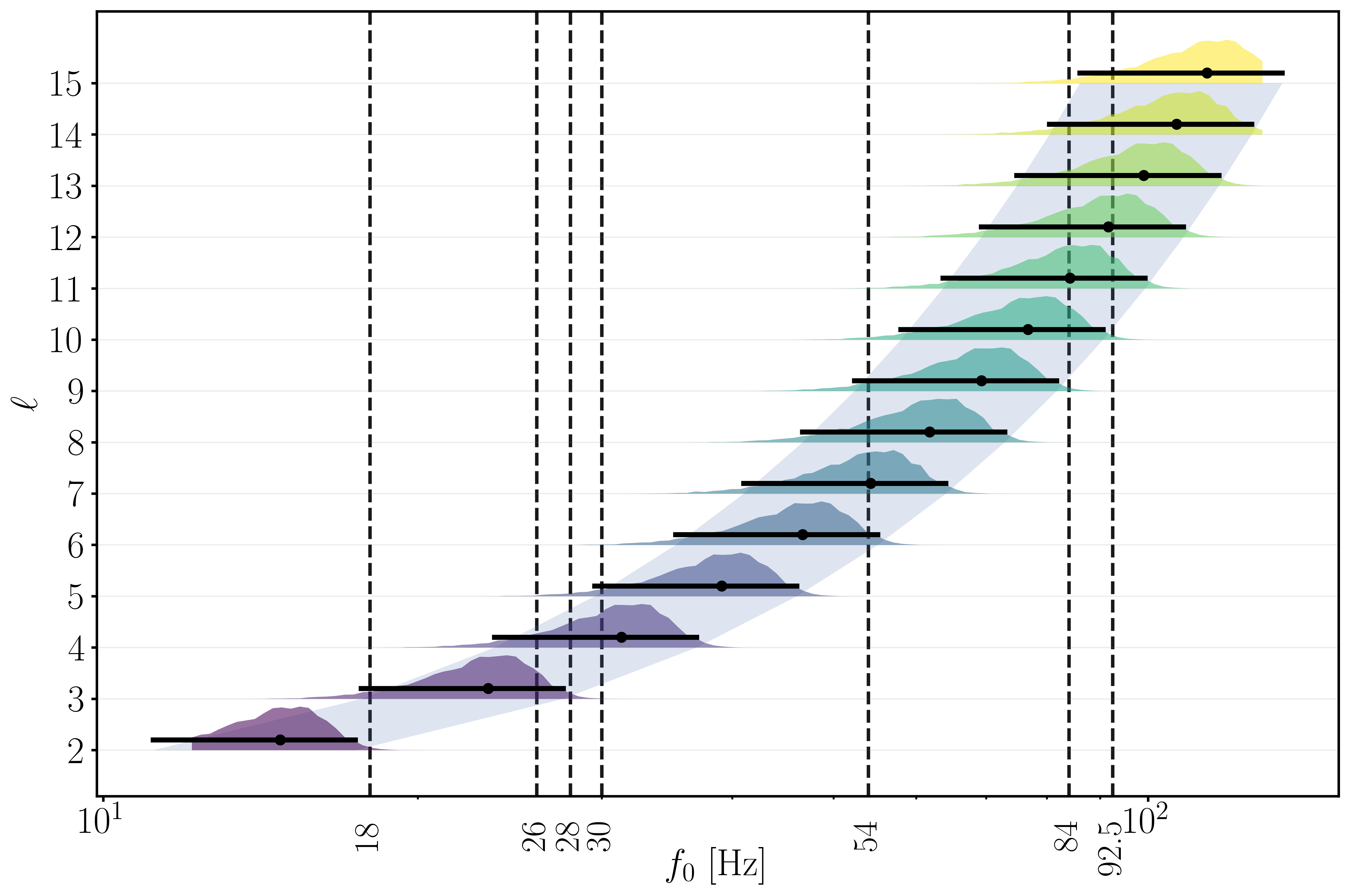}
    \caption{Posterior distributions of the fundamental crustal torsional (shear) mode frequencies
for angular indices $\ell = 2$--$15$, evaluated at a fixed neutron-star mass of
$M = 1.4\,M_\odot$. For each $\ell$, the colored ridgeline shows the normalised
frequency distribution obtained from the ensemble of equations of state, with the
black marker and horizontal bar indicating the posterior median and the corresponding
95\% credible interval, respectively. Vertical dotted lines denote the observed QPO
frequencies at 18, 26, 30, and $92.5\,\mathrm{Hz}$ in SGR~1806$-$20, and at
28, 54, and $84\,\mathrm{Hz}$ in SGR~1900+14  \cite{Watts:2006mr}.} 
    \label{fig:qpo14}
\end{figure*}

In Fig.~\ref{fig:qpo_l3_l5}, we show the posterior distributions of the fundamental
torsional mode frequencies for $\ell=3$ and $\ell=5$ as a function of neutron-star mass.
It is apparent that the observed 18~Hz QPO is compatible with the $\ell=3$ fundamental
mode across the full mass range considered, $M\simeq 1$--$2\,M_\odot$. In addition, the
30~Hz QPO can also be consistently interpreted as an $\ell=5$ fundamental mode, in
agreement with the identification proposed in Ref.~\cite{Sotani_2013}. For the 26~Hz and 28~Hz QPOs, Ref.~\cite{Sotani_2013} suggested an interpretation in
terms of the $\ell=4$ fundamental mode. However, the origin of the 26~Hz feature in
SGR~1806$-$20 remains somewhat uncertain due to its close proximity to the 30~Hz
signal.  Overall, these results indicate that, given the prevailing uncertainties in the dense-matter equation of state and crustal elastic properties, the observed low-frequency QPOs are compatible with a range of angular indices~$\ell$. At the same time, they highlight the potential of QPO observations to provide insight into the structure of the nuclear pasta layer and the neutron-star crust as more precise and abundant frequency measurements become available.

\begin{table}[t]
\caption{Posterior median values and $95\%$ credible intervals of the fundamental
crustal torsional mode frequency $f_0$ including nuclear pasta, for different neutron-star
masses.}
\label{tab:f0_pasta_ci}
\begin{ruledtabular}
\begin{tabular}{lc}
$M\;[M_\odot]$ & $f_0\;[\mathrm{Hz}]$ \\
\hline
$1.2$ & $15.28^{+2.77}_{-3.78}$ \\
$1.4$ & $14.77^{+2.66}_{-3.61}$ \\
$1.6$ & $14.29^{+2.55}_{-3.46}$ \\
$1.8$ & $13.85^{+2.46}_{-3.33}$ \\
$2.0$ & $13.51^{+2.44}_{-3.23}$ \\
\end{tabular}
\end{ruledtabular}
\end{table}

\begin{figure}
    \centering
    \includegraphics[width=1\linewidth]{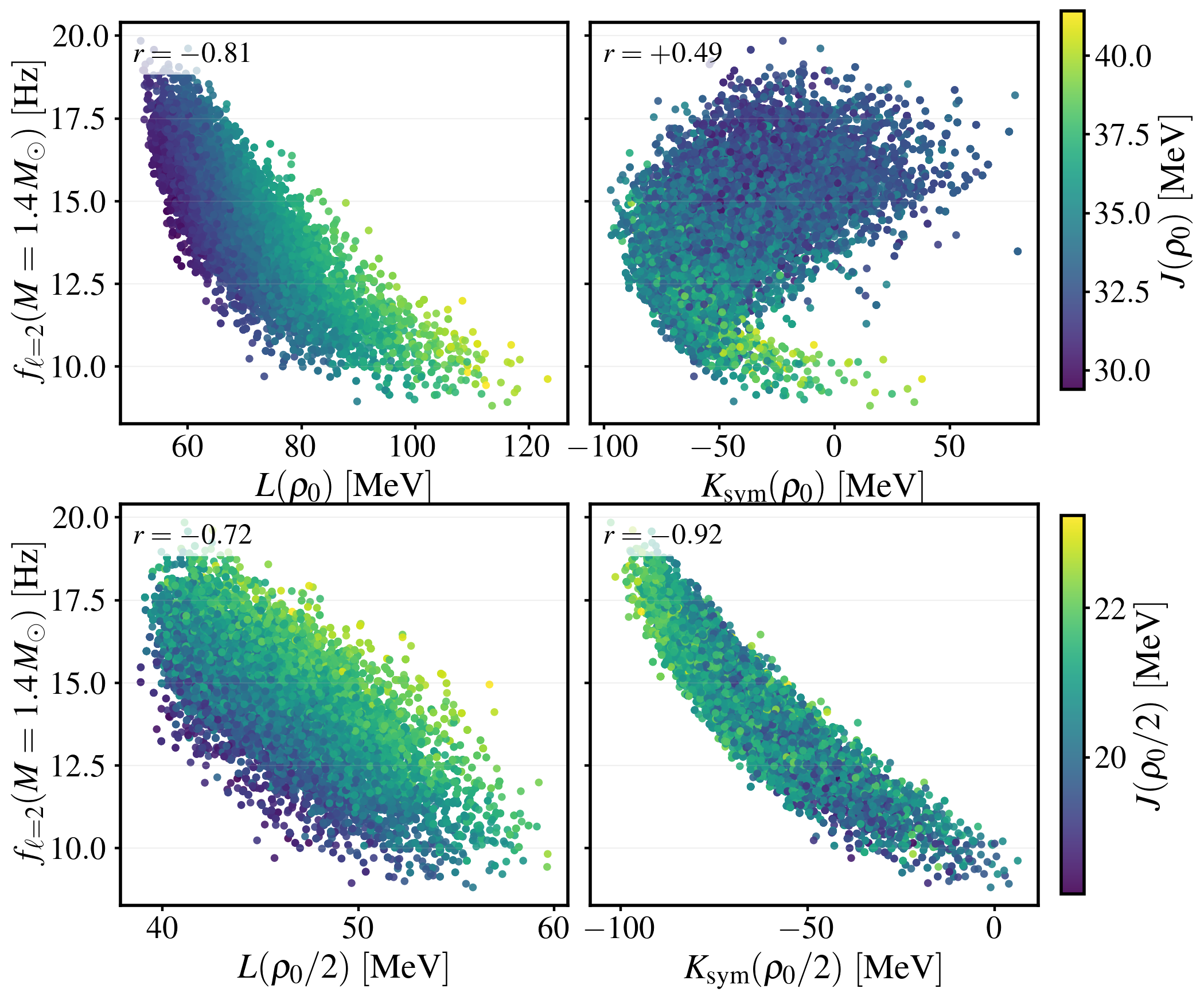}
    \caption{
Correlations between the fundamental crustal torsional (shear) mode frequency for
$\ell = 2$, evaluated at a fixed neutron-star mass of $M = 1.4\,M_\odot$, and the
nuclear-matter parameters $L$ and $K_{\rm sym}$. The upper (lower) row shows results
obtained using nuclear-matter properties evaluated at saturation density $\rho_0$
($\rho_0/2$). Points are coloured by the symmetry energy $J$ at the corresponding
density. The Pearson correlation coefficient $r$ is indicated in each panel.
}
    \label{fig:fcorr}
\end{figure}

Figure~\ref{fig:qpo_overtone} shows the posterior distribution of the first overtone
frequencies of crustal torsional  oscillations obtained from the ensemble of
equations of state, evaluated at fixed stellar mass. The results are compared with
the observed high-frequency QPO at $626.5\,\mathrm{Hz}$ in SGR~1806$-$20, which is
commonly interpreted as the first overtone of the crustal shear mode.
We find that compatibility with this frequency is achieved only for relatively
massive neutron stars. Therefore, the posterior distributions indicate that the
$626.5\,\mathrm{Hz}$ QPO cannot be explained by the $\ell=2$ first overtone and
instead requires modes with higher angular indices $\ell$.
A similar conclusion regarding the necessity of higher-$\ell$ modes for this
high-frequency QPO was previously reported in Ref.~\cite{Sotani_2011}. We support this claim based on the wide posterior distributions obtained in our analysis.

To identify the most plausible angular indices associated with the observed
quasi-periodic oscillations, we examine the joint variation of the angular index
$\ell$ and the fundamental crustal shear-mode frequency using the posterior
distributions obtained from our crust model, as shown in Fig.~\ref{fig:qpo14}.
For a given observed QPO, we associate it with a particular $\ell$ if the frequency
lies within the 95\% credible interval, indicated by the black horizontal bars, such
that the majority of the equation-of-state realizations are compatible with that
frequency. For the lowest-frequency QPO at $18\,\mathrm{Hz}$, the posterior clearly favors
$\ell = 3$. The frequencies at $26$ and $28\,\mathrm{Hz}$ are best explained by
$\ell = 4$. However, once the expected scaling of the shear-mode frequency with
$\ell$ is taken into account \cite{SteinerWatts2009PRL103, Sotani_2011, Sotani_2012, Sotani_2013}, explaining the $26\,\mathrm{Hz}$ QPO becomes
challenging within the currently allowed uncertainties of the neutron-star
equation of state. The $30\,\mathrm{Hz}$ QPO is most consistent with $\ell = 5$.
For the $54\,\mathrm{Hz}$ QPO, the posterior favors $\ell = 9$, in agreement with
the estimate $\ell = 8$ reported by Ref.~\cite{Sotani_2013}. In this case, both
$\ell = 8$ and $\ell = 9$ provide viable explanations, with $\ell = 9$ being more
strongly favored by the posterior. Similarly, the $84\,\mathrm{Hz}$ QPO can be explained by both $\ell = 13$ and
$\ell = 14$, with a preference for $\ell = 14$, while the $92.5\,\mathrm{Hz}$ QPO
is most naturally associated with $\ell = 15$. These identifications are summarized
in Table~\ref{tab:qpo_l_match}, which provides an updated mapping between the observed
QPO frequencies and the corresponding angular indices $\ell$ based on current
equation-of-state constraints. We note that Ref.~\cite{Gearheart_2011} interpreted the high-frequency QPOs at
$84$ and $92.5\,\mathrm{Hz}$ as first overtones for very large values of the symmetry
energy slope parameter $L$. Such large values of $L$ are now strongly disfavored by
modern nuclear and astrophysical constraints, and we therefore do not find support
for this interpretation in the present analysis. Finally, we observe that the
preferred angular index $\ell$ increases approximately exponentially with the QPO
frequency. In the present work, we have not included neutron–drip entrainment effects \cite{Gearheart_2011}.
The inclusion of entrainment is expected to reduce the crustal shear-mode
frequencies, and the values reported here should therefore be interpreted as
upper limits on the corresponding angular indices~$\ell$.

\begin{table}[t]
\centering
\caption{Best-matching angular indices $\ell$ for the observed QPO frequencies
based on the posterior distributions of fundamental crustal shear-mode frequencies
at fixed mass $M=1.4\,M_\odot$. The quoted $\ell$ values correspond to those for which
the observed frequency lies within the 95\% credible interval and is supported by
the majority of equation-of-state realisations.}
\label{tab:qpo_l_match}
\begin{tabular}{ccc}
\hline\hline
Source & QPO frequency [Hz] & Favoured $\ell$ \\
\hline
SGR~1806$-$20 & 18   & 3 \\
              & 26   &  \\
              & 30   & 5 \\
              & 92.5 & 15 \\
\hline
SGR~1900+14   & 28   & 4 \\
              & 54   & 9 \\
              & 84   & 14 \\
\hline\hline
\end{tabular}
\end{table}

To explore how the fundamental crustal torsional (shear) mode frequency depends on nuclear-matter properties, we examine its correlation with the symmetry-energy slope $L$ and curvature $K_{\rm sym}$ for a neutron star of mass $M = 1.4,M_\odot$. The scatter plots shown in Fig.~\ref{fig:fcorr} display the resulting distributions when these parameters are evaluated at saturation density $\rho_0$ and at sub-saturation density $\rho_0/2$. As expected for crust-dominated observables, the correlations are significantly stronger when $L$ and $K_{\rm sym}$ are taken at $\rho_0/2$, reflecting the dominant role of the symmetry energy in the density regime relevant for the neutron-star crust.
 We find that, similar to other crustal properties, the QPO frequency
exhibits a significantly stronger correlation with both $L$ and $K_{\rm sym}$ when
these quantities are evaluated at $\rho_0/2$. In particular, $K_{\rm sym}$ shows the
strongest correlation with the fundamental QPO frequency among the considered
parameters. We have verified that this behaviour persists for higher angular indices, with the
overall structure and relative strength of the correlations remaining qualitatively
unchanged. When combined with the trend observed in Fig.~\ref{fig:qpo14}, where the
fundamental mode frequency scales approximately exponentially with the angular index
$\ell$, these results indicate that future, more precise measurements of QPO
frequencies can provide robust constraints on the density dependence of the symmetry
energy, especially through $K_{\rm sym}$ evaluated at subsaturation densities.

\section{\label{conclusion} Conclusion}

In this work we investigate the impact of nuclear pasta on neutron-star crust
properties and crustal torsional oscillations within a unified and statistical framework. Starting from a Bayesian posterior ensemble of relativistic
mean-field equations of state constrained by nuclear experiments, chiral effective
field theory at sub-saturation densities, and multimessenger astrophysical
observations, we computed the inner-crust structure and pasta phases for each
posterior sample using a compressible liquid-drop model. This approach allowed us
to quantify uncertainties in the onset, extent, and mass fraction of nuclear pasta
in a consistent manner.

We find that the transition from spherical nuclei to cylindrical rods is 
well constrained. Beyond the rod phase, the pasta sequence is more
model dependent. While all equations of state predict at least one non-spherical
phase, the majority proceed directly from rods to uniform matter, with only a
fraction of models supporting an additional slab phase and a negligible subset
reaching more complex geometries. When extended pasta structures are present, the
crust--core transition is shifted to slightly higher densities, reflecting the
energetic favorability of non-spherical configurations close to the transition.
We further showed that the appearance and extent of pasta are strongly regulated by
the isovector sector of the equation of state at sub-saturation density, with lower
values of the symmetry-energy slope and more negative curvature favoring extended
pasta layers.

Using the resulting crust models, we computed crustal torsional shear-mode
frequencies in a plane-parallel approximation, incorporating a smooth suppression
of the shear modulus in the pasta region and enforcing the correct spherical-mode
scaling in the non-magnetic limit. We find that the inclusion of nuclear pasta
systematically reduces the crustal shear modulus near the crust--core interface and
shifts the fundamental torsional frequencies to lower values. Within our posterior
ensemble, the $\ell=2$ fundamental mode cannot account for the observed
$18~\mathrm{Hz}$ QPO in SGR~1806$-$20 once pasta is included, consistent with earlier
studies. Instead, the observed low-frequency QPOs are compatible with higher-$\ell$
fundamental modes, and we find that, given current equation-of-state uncertainties,
multiple angular indices can provide viable interpretations.

We also quantified correlations between crustal oscillation frequencies and
nuclear-matter parameters. The strongest correlations are obtained when the symmetry
energy parameters are evaluated at sub-saturation density, with
$K_{\mathrm{sym}}(\rho_0/2)$ showing a particularly strong influence on the
fundamental torsional frequencies. This highlights the potential of crustal QPOs,
when combined with a consistent treatment of the inner crust, as indirect probes of
the density dependence of the symmetry energy below saturation.

The present analysis isolates the role of nuclear pasta in a non-magnetic framework
and neglects entrainment and superfluid effects, which are expected to further
reduce crustal shear-mode frequencies. Including these effects, together with
full magneto-elastic coupling to the core, which will be important for a complete
interpretation of magnetar QPOs. Nevertheless, our results demonstrate that nuclear
pasta produces  softening of the inner crust and leads to measurable
shifts in the torsional mode spectrum within the currently allowed dense-matter
uncertainties.

\bibliography{pasta_bays}
\end{document}